\documentstyle[twocolumn,prl,aps]{revtex}
\begin{document}
\draft
\title{Pionic Measures of Parity and $\cal CP$ Violation 
in High Energy Nuclear Collisions}
\author{Dmitri Kharzeev$^a$ and Robert D. Pisarski$^b$}
\bigskip
\address{
a) RIKEN BNL Research Center, Brookhaven National Laboratory,
Upton, New York 11973-5000, USA\\
email: kharzeev@bnl.gov\\
b) Department of Physics, Brookhaven National Laboratory,
Upton, New York 11973-5000, USA\\
email: pisarski@bnl.gov\\
}
\date{\today}
\maketitle
\begin{abstract}
The collisions of large nuclei at high energies could produce metastable
vacua which are odd under parity, $\cal P$, charge
conjugation, $\cal C$, and/or 
$\cal CP$.  Using only the three-momenta of charged
pions (or kaons), we show how to construct global observables which are odd
under $\cal P$, $\cal C$, and
$\cal CP$, and which can be measured on an
event by event basis.  Model dependent estimates 
of the $\cal P$-odd observables are on the order of $10^{-3}$.
\end{abstract}
\pacs{}

In a previous Letter we proposed a detailed model in
which metastable vacua could arise in hot $QCD$\cite{pap}.  
These metastable
states are related to $\theta$-vacua, and so are odd
under charge conjugation times parity, $\cal CP$.  
(For related studies, especially in supersymmetric theories,
see \cite{shifman,wittennew,zhitnitsky}.)
An obvious question
is how the breaking of this discrete symmetry by a metastable bubble
could be measured in the collisions of large nuclei at high energies.
As the bubble is odd under $\cal P$ and $\cal CP$,
the pions produced by its decay must also be in a state which is
odd under these symmetries. In \cite{pap} we
proposed measuring, on an event by event basis, 
a global variable which is odd under $\cal P$.

In this article we do two things.  
Using only the momenta of charged pions in an event, we construct
global observables which are odd under the discrete symmetries of 
$\cal P$, $\cal C$, and/or $\cal CP$.
Our discussion is general, 
independent of whatever detailed mechanism might produce nonzero 
values for these variables.  For the collisions of nuclei with equal atomic
number, as the initial state is even under $\cal P$, the observation
of a $\cal P$-odd final state must be due to parity violation, such
as by a $\cal P$-odd bubble.  
Based upon our specific model \cite{pap}, we then give a rough estimate of 
the magnitude of the $\cal P$-odd and $\cal CP$-odd effects;
we find that the asymmetries can be relatively
large, at least $\sim 10^{-3}$.

At high energy, nucleus-nucleus collisions produce many pions,
on the order of $\sim 1000$ per unit rapidity at RHIC energies.
Experimentally, it is probably easiest to detect charged pions and their
three-momenta.  (All of our comments apply equally well to charged
kaons.)  Thus we are led to consider constructing observables
only from the three-momenta of $\pi^+$'s, $\vec{p}_+$, and 
$\pi^-$'s, $\vec{p}_-$.  As vectors, under parity the three-momenta 
transform as 
\begin{equation}
{\cal P}: \;\;\; \vec{p}_+ \rightarrow - \vec{p}_+ \;\;\; , \;\;\; 
\vec{p}_- \rightarrow - \vec{p}_- \; .
\end{equation}
Charge conjugation switches 
$\pi^+$ and $\pi^-$, 
\begin{equation}
{\cal C}: \;\;\; \vec{p}_+ \leftrightarrow \vec{p}_-  .
\end{equation}
It is a theorem that any $\cal P$-odd invariant 
formed from three-vectors can be represented as a sum of terms,
each of which involves one antisymmetric epsilon tensor \cite{weyl}.
The variable which we proposed previously is of this type \cite{pap}:
\begin{equation}
{\bf J} = 
\sum_{\pi^+, \pi^-}
(\hat{p}_{+} \times \hat{p}_{-})\cdot \hat{z} \; .
\label{eqj}
\end{equation}
In order to form ${\bf J}$ we have introduced an arbitrary, fixed
vector of unit norm, $\hat{z}$.  If $\hat{z} \rightarrow - \hat{z}$
under parity, then ${\bf J}$ is odd under ${\cal P}$.
In ${\bf J}$ we use the unit vectors 
$\hat{p}_\pm = \vec{p}_\pm/|\vec{p}_\pm|$ so that
it is a pure, dimensionless number.
The variable ${\bf J}$ is separately odd under $\cal P$ and $\cal C$,
and so is even under $\cal CP$.  

The variable $\bf J$ is closely analogous to ``handedness'', originally
introduced to study spin dependent effects in jet
fragmentation \cite{hand1}.
There the axis $\hat{z}$ is usually defined by
the thrust of the jet, with $\hat{p}_+$ and $\hat{p}_-$ representing
the directions of pions formed in the fragmentation of the jet.
Correlations between the handedness of different jets produced
in a given event are sensitive to $\cal CP$-violating effects \cite{hand2}.

It is not difficult to construct
other invariants with different transformation properties.  We
introduce $\vec{k}_\pm$ as
\begin{equation}
\vec{k}_\pm = \sum_{\pi^+} \vec{p}_+ \pm \sum_{\pi^-}\vec{p}_- \;\;\; , \;\;\;
\hat{k}_\pm = \vec{k}_\pm/k_\pm \; , 
\end{equation}
and then form
\begin{equation}
{\bf K}_\pm = 
\sum_{\pi^+, \pi^-}
(\hat{p}_{+} \times \hat{p}_{-})\cdot \hat{k}_\pm \; .
\label{eqk}
\end{equation}
The variables ${\bf K}_\pm$ are $\cal P$-odd;
${\bf K}_+$ is $\cal C$-odd, and so $\cal CP$-even,
while ${\bf K}_-$ is $\cal C$-even, and so $\cal CP$-odd.
The vector $\vec{k}_+$ measures the net flow of the charged pion momentum,
while $\vec{k}_-$ measures the net flow of charge from pions.

We can also form
\begin{equation}
{\bf L} = 
\sum_{\pi^+, \pi^-}
(\hat{p}_{+} \times \hat{p}_{-})\cdot \hat{z} 
\; \; \left( \frac{p_+^2 - p_-^2}{p_+^2 + p_-^2} \right) \; .
\end{equation}
The variable $\bf L$ is $\cal P$-odd,
$\cal C$-even, and $\cal CP$-odd.  It does not require a net flow
of momentum or charge to be nonzero, although as for $\bf J$, we do need
to introduce an arbitrary unit vector $\hat{z}$.

Similar to $\bf L$, we can form \cite{altesun}
\begin{equation}
{\bf M} = 
\sum_{\pi^+, \pi^-}
\; \; \left( \frac{p_+^2 - p_-^2}{p_+^2 + p_-^2} \right) \; .
\end{equation}
This variable is $\cal P$-even, $\cal C$-odd, and so $\cal CP$-odd.

Besides using the vectors $\vec{k}_\pm$, another way of avoiding
the introduction of an arbitrary unit vector $\hat{z}$ is to 
use ordered pairs of pion momenta \cite{lee,gyulassy};
this procedure is also used in the studies of spin-dependent
effects in jet fragmentation \cite{hand1,hand2}.  For any given
pair of like sign pions, let $\vec{p}^{\; \prime}_+$ denote the $\pi^+$ with
largest momentum, so $|\vec{p}^{\; \prime}_+| > |\vec{p}_+|$.  This ordering
is done in the frame in which the observable is measured.
Then we can form a triple product as
\begin{equation}
{\bf T}_\pm = 
\sum_{\pi^+, \pi^-}
(\hat{p}_{+} \times \hat{p}_{-})\cdot (\hat{p}^{\, \prime}_{+} 
\pm \hat{p}^{\, \prime}_{-}) \;\;\; ,
\;\;\; |\vec{p}^{\; \prime}_\pm| > |\vec{p}_\pm|
\end{equation}
The variables ${\bf T}_\pm$ are $\cal P$-odd;
${\bf T}_+$ is $\cal C$-odd and $\cal CP$-even,
while ${\bf T}_-$ is $\cal C$-even and $\cal CP$-odd.
Besides the variables ${\bf T}_\pm$, one can clearly construct
other $\cal P$- and $\cal C$-odd observables from triplets,
or higher numbers, of pions.

The metastable $\cal P$-odd bubbles of ref. \cite{pap} were derived
within the context of a nonlinear sigma model, with 
a $U(3)$ matrix $U$, $U^\dagger U = \bf 1$.  
The metastable vacua are stationary points with respect to the
nonlinear sigma model action, including the terms with two derivatives,
a mass term, and an anomaly term \cite{effective}:
$$
{\cal L} = f_\pi^2 \left\{ \frac{1}{2} \; 
tr\left(\partial_\mu U^\dagger \partial_\mu U \right) \; + \;
c \; tr\left( M(U + U^\dagger) \right) \right.
$$
\begin{equation}
\left. - a (tr \; ln \; U - \theta)^2 \right\}\; ,
\end{equation}
where $f_\pi = 93 MeV$ is the pion decay constant,
and $\theta$ is the $\theta$-parameter.  In $QCD$,
$\theta \leq 10^{-9}$; we retain $\theta$ for now, because
it helps illuminate the nature of the $\cal P$-odd bubbles.
$M$ is the mass matrix for current quark masses, 
$M = diag(m_1, m_2,m_3)$, and  $c$ is a constant.
In this paper we only need the ratios of these quantities:
taking $m_1 = m_u$, $m_2 = m_d$, $m_3 = m_s$,
for the up, down, and strange quark masses, respectively, 
we take $m_u/m_d \approx 1/2$ and $m_u/m_s \approx 1/36$.
The anomaly term
is proportional to the topological susceptibility,
$a \sim \int d^4 x \, \langle Q(x) Q(0)\rangle$,
where $Q$ is the topological charge density,
$Q= g^2/(32 \pi^2) tr(G_{\mu \nu} \widetilde{G}^{\mu \nu})$ \cite{effective}.
At zero temperature, $a$ is large, and the $\eta'$ meson is heavy,
$m^2_{\eta'} \sim a$.  

We are interested in stationary points of the effective lagrangian,
${\cal L}$.  In our previous work \cite{pap}, for simplicity we assumed that
the stationary point is 
constant in space and time, so that only the mass and anomaly terms
enter into the equations of motion.  
By a global chiral rotation,
a constant $U$ field can be rotated into a diagonal matrix.  With
$U_{i j} = exp(i \phi_i) \;\delta^{i j}$, the effective potential
becomes
\begin{equation}
V(\phi_i) = f_\pi^2 \left(
- c \sum_{i} m_i \; cos(\phi_i)
+ \frac{a}{2} (\sum_i \phi_i - \theta )^2 \right) \; ,
\label{e3}
\end{equation}
At a stationary point, the $\phi_i$'s satisfy
\begin{equation}
c \; m_i \; sin(\phi_i) \; =  \; a (\phi_1 + \phi_2 + \phi_3 -\theta ) \; .
\end{equation}
We are interested in solutions for which the $\phi_i \neq 0$.
It is clear from the stationary point condition that nonzero 
values of $\phi_1 + \phi_2 + \phi_3$ act like having a system with
$\theta \neq 0$.  With this insight, we set $\theta = 0 $ henceforth.

As the anomaly term in (\ref{e3}) arises from $tr \, ln \, U$, 
$\sum_i \phi_i$ is defined to be periodic modulo $2 \pi$.
Consequently, when the anomaly term vanishes,
$a=0$, then any $\phi_i$ equal to a multiple of $2 \pi$
is a solution, but by periodicity these are all equivalent to the trivial
solution, $\phi _i = 0$.  When $a \neq 0$, however, Witten
observed \cite{wittenold} that there may be 
nontrivial solutions, in which some
$\phi_i$ are {\it near} a multiple of $2 \pi$.  These are not equivalent
to the trivial vacuum, and represent the spontaneous breaking of
$\cal P$ and $\cal CP$ symmetries, in exactly the same way as
$\theta \neq 0$ violates $\cal CP$ symmetry.  These solutions only arise
when $a$ is sufficiently small.
Based upon an analysis in the limit
of a large number of colors, we suggested 
that near the phase transition, $a$ becomes much smaller than
its value at zero temperature \cite{pap}.  
In a mean field type of analysis, with $T_d$ the temperature
of the deconfining transition, and $t = (T_d -T)/T_d$ the
reduced temperature, we found that 
$c(T) \sim 1/t^{1/2}$ and
$a(T) \sim t$, so that the relevant ratio, $a/c$, scales
as $a(T)/c(T) \sim t^{3/2}$.  
With the parameters of our model, we find that 
$\cal P$-odd bubbles 
appear once $a/c$ is $\sim 1\%$ of its value
at zero temperature.  As $a/c \rightarrow 0$, these
solutions satisfy 
$\phi_1 \approx 2 \pi - \phi_u$, 
$\phi_2 \approx - \phi_d$, and $\phi_3 \approx - \phi_s$,
where $m_u \phi_u \approx m_d \phi_d \approx m_s \phi_s$.
For example, in our model, $\cal P$-odd bubbles first appear
when $a/c < (a/c)_{cr} \sim .24$.  At this point,
$\phi_u \approx 1.8$, $\phi_d \approx .5$, and
$\phi_s \approx .03$.  We stress that this is only the stationary point
with minimal action; for arbitrary integers $n = \pm 1, \pm 2 \ldots$, 
there exist other stationary points with 
$\phi_1 \approx 2 n \pi$, and
$\phi_2 \approx \phi_3 \approx 0$, with energy densities which grow like
$\sim n^2$.  

In terms of the underlying gluonic fields, the $\cal P$-odd bubbles
arise from fluctuations in the topological charge density, $G_{\mu \nu}
\widetilde{G}^{\mu \nu}$.  It is easy to understand how
a region in which $G_{\mu \nu} \widetilde{G}^{\mu \nu} \neq 0$
can produce a $\cal P$-odd effect.  Consider the propagation of
a quark anti-quark pair through a regioin in which 
$G_{\mu \nu} \widetilde{G}^{\mu \nu} \neq 0$; in terms of
the color electric, $\vec{E}$, and color magnetic, $\vec{B}$,
fields, $G_{\mu \nu} \widetilde{G}^{\mu \nu} \sim \vec{E} \cdot \vec{B}$.
If $\vec{E}$ and $\vec{B}$ both lie along the $\hat{z}$ direction,
then a quark is bent one way, the anti-quark the other,
so that $(\vec{p}_q \times \vec{p}_{\overline{q}}) \cdot \hat{z} \neq 0$,
where $\vec{p}_q$ and $\vec{p}_{\overline{q}}$ are the three-momenta
of the quark and anti-quark, respectively.  While physically
intuitive, this picture does not allow us to directly relate
this bending in the momenta of the quark and anti-quark to an asymmetry
for charged pions.  

To do so, we again resort to using an effective lagrangian.
It is known that the effects of the axial anomaly show up
in the effective lagrangians of Goldstone bosons in two 
\cite{effective,wittenold,wz,wittenwz},
and only two \cite{dhoker}, ways.
One is through the anomaly
term \cite{effective,wittenold}, 
$\sim a$, which we have already included.  Besides that,
however, the axial anomaly also manifests itself in the interactions of
Goldstone bosons through the Wess-Zumino-Witten term \cite{wz,wittenwz}.
This term is nonzero only when the fields are time dependent,
which is why we could ignore it in discussing the static properties
of $\cal P$-odd bubbles.  It cannot be ignored, however, in 
computing the dynamical properties, and in particular the decay,
of $\cal P$-odd bubbles.
The Wess-Zumino-Witten term is manifestly chirally symmetric
when written as an integral over five dimensions,
\begin{equation}
{\cal S}_{wzw} = 
- i {1\over 80 \pi^2} \int d^5 \!x \;
\varepsilon^{\alpha \beta \gamma \delta \sigma}\,
tr\left(R_\alpha R_\beta R_\gamma R_\delta R_\sigma 
\right ) \; , 
\end{equation}
$R_\alpha = U^\dagger \partial_\alpha U $,
but reduces to a boundary 
term in four space-time dimensions.  For $U = exp( i u)$,
when $\partial_\alpha u \ll 1$,
\begin{equation}
{\cal S}_{wzw} \approx
{2\over 5 \pi^2} \int d^4 \!x \;
\varepsilon^{\alpha \beta \gamma \delta }\,
tr\left(u \; \partial_\alpha u \; 
\partial_\beta u \; \partial_\gamma u \;
\partial_\delta u
\right ) \; .
\label{wzw1}
\end{equation}
As discussed by Witten \cite{wittenwz}, the coefficient of
the Wess-Zumino-Witten term is fixed by topological considerations,
and is
proportional to the number of colors, which equals three.

In a collision, we envision that the trivial vacuum heats up,
a $\cal P$-odd bubble forms, and then decays as the vacuum cools.
Since this represents bubble formation and decay, there is no
net change in any topological number.  Therefore, it is possible for
a given event to contain an excess of bubbles over anti-bubbles
(or vice versa), and thus to manifest true parity violation on an
event by event basis.

To estimate the magnitude of the Wess-Zumino-Witten term for
a $\cal P$-odd bubble, and to understand its effect on pion
production, in (\ref{wzw1}) we 
can take three $u$'s to be condensate fields, $u \sim \phi_{u,d,s}$,
and two to be charged pion fields, $u \sim \pi_{\pm}/f_\pi$.  
Suppose that the 
$\cal P$-odd bubble is of size $R$, with unit normal
$\hat{r}$ to the surface, and lasts for some period of time.
Because of the antisymmetric tensor in
(\ref{wzw1}), all three components of the condensate field
must enter.  Schematically, we obtain
\begin{equation}
{\cal S}_{wzw} \approx
{2\over 5 \pi^2} \;
\int dt \int d^3 r  \; \phi_u \; \partial_r \phi_d \; \partial_0 \phi_s \;
(\vec{p}_{\pi^+} \times \vec{p}_{\pi^-}) \cdot \hat{r}
\end{equation}
The time integral is 
$\int dt \, \partial_0 \phi_s \sim \delta \phi_s = \phi_s$, since 
$\phi_s = 0$ in the
normal vacuum.  Similarly, the spatial integral is
$\int d^3r \, \partial_r \phi_d \, (\vec{p}_{\pi^+} \times \vec{p}_{\pi^-}) 
\cdot \hat{r}
\sim \int d\Omega \, \int R^2 dr \, \partial_r \phi_d \,
(\vec{p}_{\pi^+} \times \vec{p}_{\pi^-}) 
\cdot \hat{r} 
\sim \int d\Omega \, R^2 \phi_d \,
(\vec{p}_{\pi^+} \times \vec{p}_{\pi^-}) 
\cdot \hat{r}$, where $\int d\Omega$ represents an
integral over the direction of the normal, $\hat{r}$.  
Further, as the average momentum within the condensate is
of order $p_\pi \sim 1/R$, the size of the bubble drops out as well.
We thus obtain a final result which is independent of the size
of the bubble, its lifetime, and its width:
\begin{equation}
{\cal S}_{wzw} \approx
\frac{2 \phi_u \phi_d \phi_s}{5 \pi^2} \int d \Omega \;
(\hat{p}_{\pi^+} \times \hat{p}_{\pi^-}) \cdot \hat{r}
\label{wzw2}
\end{equation}
We stress that it is only the decay of the $\cal P$-odd
bubble which produces an effect, since the Wess-Zumino-Witten term
vanishes for a static field.  Also, note that there is only
an observable asymmetry when $\phi_s \neq 0$; this is because
in the absence
 of external gauge fields, there is only a Wess-Zumino-Witten
term for three, and not for two, flavors.  Within this model,
${\cal S}_{wzw}$ is of similar form for two charged kaons.

Using our previous estimates for the $\phi$'s, $\phi_u \sim \phi_d \sim 1$
and $\phi_s \sim 10^{-2}$, we obtain an effect of order
$\sim 10^{-3}$.
At the point where the $\cal P$-odd bubble first appears,
$(a/c)_{cr}$, one can estimate that the energy density
within the bubble, relative to the ordinary
vacuum, is $\sim 25 \, n^2 \, MeV/fm^3$, where
$n$ is the winding number of the bubble, $n=1,2,3...$
For a bubble $\sim 5 fm$ in radius, there are about
$\sim 100 \, n^2$ pions produced in the decay of the bubble.
If a fraction of the produced pions are observed within a
given kinematical window, and we assume that 
all observed pions come from a portion of the total bubble,
then we recover the variable $\bf J$, introduced before in
(\ref{eqj}), and find an estimate of ${\bf J} \sim 10^{-3}$.  
Moreover, we find a natural interpretation of the direction
$\hat{z}$, which was needed to define $\bf J$, as the normal
to the bubble's surface.  One might wonder if the effect is
diluted by the necessity to average over uncorrelated pairs.
This does not happen, however, because the pion field within
the bubble is a classical field, so that all charged pions are
affected similarly.  

Naively, one might expect that $\bf J$ would average to zero over
a single bubble.  As the bubble is topological, though, the
direction in which charged pions are swept is correlated with
the sign of the condensate, so that a single $\cal P$-odd bubble
can produce an effect in ${\bf J} \sim 10^{-3}$.  
Thus it is possible to distinguish between events in which bubbles are
produced, and those in which bubbles are not, by measuring
$\bf J$.

At first it may seem surprising that our $\cal P$-odd,
$\cal C$-even, and $\cal CP$-odd bubble
produces a signal in $\bf J$, which by previous analysis is
$\cal P$-odd, $\cal C$-odd, and $\cal CP$-even.  
The Wess-Zumino-Witten term is even under parity, which
is ${\cal P}_0 (-1)^{N_B}$, where ${\cal P}_0$ is the operation
of space reflection, and $N_B$ counts the number of 
Goldstone bosons \cite{wittenwz}.  By
scattering off a $\cal P$-odd bubble, we bring in
an odd number of condensate fields, ${\bf J} \sim \phi_u \phi_d \phi_s$,
(\ref{wzw2}), 
and so change the (apparent) quantum numbers to be $\cal P$-odd and
$\cal CP$-even.
This is only apparent, as scattering off an anti-bubble will give
the opposite sign of $\bf J$.

We expect that bubbles will generate signals for the other
variables presented of similar magnitude.  For example, a
single bubble will induce a 
net flow of pion charge, and so contribute to
${\bf K}_- \sim 10^{-3}$, (\ref{eqk}).  Through coherent
scattering in a bubble, we would also expect the variables
${\bf K}_+$, $\bf L$, and $T_\pm$ to develop signals
$\sim 10^{-3}$. Further, hot gauge
theories can also exhibit metastable states which are 
$\cal P$-even and $\cal C$-odd
\cite{altes}; these generate signals for $\bf M$ \cite{altesun}.

The idea of exciting metastable vacua in hadronic reactions
is an old one \cite{leewick}, as is the idea that
a collective pion field can produce large fluctuations
in heavy ion collisions on an event by event basis
\cite{dcc}.  We wish to emphasize that there are
certain topologically nontrivial configurations of pion fields
which produce signals that are
odd under the discrete symmetries of $\cal P$, $\cal C$, and/or
$\cal CP$.  Any observation of such
violation of these discrete symmetries
would be, prima facie, evidence of novel physics.
Whatever credence one ascribes to our detailed dynamical model,
the observables which we propose herein are possible
to measure \cite{sand}, and we strongly encourage our experimental
colleagues to do so.

We wish to thank M. Gyulassy, C. P. Korthals Altes, 
T.~D.~Lee, J. Sandweiss, and E. V. Shuryak for discussions.
The work of R.D.P. is supported by a DOE grant at 
Brookhaven National Laboratory, DE-AC02-76CH00016.

\end{document}